\documentclass[epjST]{svjour}
\usepackage{graphics}
\usepackage{amsmath}
\usepackage{amsfonts}
\usepackage{amssymb}
\usepackage{xcolor}
\def\beq{\begin{equation}}
\def\eeq{\end{equation}}
\def\bes{\begin{subequations}}
\def\ees{\end{subequations}}
\def\bea{\begin{eqnarray}}
\def\eea{\end{eqnarray}}
\def\bry{\begin{array}}
\def\ery{\end{array}}
\def\bit{\begin{itemize}}
\def\eit{\end{itemize}}
\def\nn{\nonumber}

\def\tr{\textrm{Tr}}

\def\t{\widetilde}
\def\dd{\displaystyle}

\def\dst{\displaystyle}
\def\f{\frac}
\def\gst{g_{V}}
\def\eq{Eq.~\eqref}
\def\nn{\nonumber}

\def\lmu{{\bf L}_\mu}
\def\rmu{{\bf R}_\mu}

\def\gs{{g''}}
\def\Wt{\tilde{\bf {W}}}

\def\Yt{\tilde{\bf {Y}}}

\def\gt{\tilde g}
\def\gpt{{{\tilde g}^\prime}}
\def\de{\partial}
\def\rhot{\tilde\rho}
\def\vv{\dd{v^2}}
\def\sf{s_{\varphi}}
\def\cf{c_{\varphi}}
\begin{document}
\title{Spin-1 resonances}
\author{Stefania De Curtis\inst{1}\fnmsep\thanks{\email{decurtis@fi.infn.it}} \and Daniele Dominici\inst{1} \fnmsep\thanks{\email{dominici@fi.infn.it}}  }
\institute{INFN, Sezione di Firenze and Department of Physics and Astronomy,
 University of Florence\\Via G. Sansone 1, 50019 Sesto Fiorentino, Italy 
}
\abstract{Spin-1 resonances are naturally present in composite Higgs frameworks. We first review  a model independent approach to parameterize a single additional heavy triplet  and then we consider more realistic models arising in  composite Higgs scenarios where a larger number of spin-1 resonances is expected. In these cases, finite width and interference effects can heavily affect the bounds  extracted from the data.
} 
\maketitle
\section{Introduction}
\label{intro}
The presence of additional spin-1 particles, triplets or singlets of $SU(2)$, is a common feature of several beyond the Standard Model  (BSM) frameworks. They appear as Kaluza-Klein excitations of $W$, $Z$ and $\gamma$ in flat or warped extra-dimension extensions of the SM, as new vector and axial-vector resonances in walking technicolor inspired effective Lagrangians, and are naturally present in composite Higgs scenarios (for references see, for example, the recent reviews \cite{Panico:2015jxa,Cacciapaglia:2020kgq}). They are also expected in weakly interacting BSM theories like the Little Higgs models or Grand Unified theories. In this chapter we review some of the topics related to the theoretical approaches to describe the interactions of the new spin-1 resonances with the SM particles and with a composite Higgs. The new vectors could give clear signals in the Drell-Yan channel at the LHC, in the dilepton or in the diboson channels. From the LHC measurements one could then derive bounds on their masses and  couplings.
A very convenient tool to get these limits is a general model independent approach \cite{Pappadopulo:2014qza}. This allows, from a comparison with the data, to get bounds on the couplings to fermions and  bosons of a new spin-1 triplet  at fixed mass.
This simplified approach however is not able to reproduce extended models containing more than one triplet. We consider then possible proposals describing two new triplets which can describe the low-lying resonances relevant at the LHC and outline the importance  of the interference and finite width effects.

In Section \ref{sec:2} we review a model independent approach, to parameterize the presence of an additional heavy triplet of spin-1 resonances, that makes use of an effective Lagrangian description.  Experimental data from CMS and ATLAS collaborations, using this parameterization, are used in Section \ref{sec:3} to get bounds on two particular models. In Section \ref{sec:4} we review a general method to build effective Lagrangians describing two new heavy triplets, replicas of $W$ and $Z$,  analize two particular models and review the bounds on their parameter space  using the LHC measurements. Finally in Section \ref{sec:5} we outline, by considering a composite Higgs model, the relevance of the interference and finite width effects in the analysis of the data to extract bounds on the model itself.

\section{Heavy vector triplet simplified model and bounds from the LHC measurements}
\label{sec:2}
Following \cite{Pappadopulo:2014qza} (see also  \cite{deBlas:2012qp}) we can parameterize the presence of a new heavy vector triplet, interacting with SM gauge bosons and fermions, by the following simple effective phenomenological Lagrangian containing operators up to dimension $d=4$
\bea
\dd{\mathcal{L}}_V &=&\dd-\frac 1 4 D_{[\mu}V_{\nu ]}^a D^{[\mu}V^{\nu ]\;a}+\frac{m_V^{2}}2V_\mu^a V^{\mu\;a}
\dd 
+\, i\,\gst  c_H V_\mu^a H^\dagger \tau^a {\overset{{}_{\leftrightarrow}}{D}}^\mu H+\frac{g^2}{\gst } c_F V_\mu^a J^{\mu\;a}_F\vspace{2mm}\nn\\
&&\dd+\frac{\gst }2 c_{VVV} \epsilon_{abc}V_\mu^a V_\nu^b D^{[\mu}V^{\nu] c}+\gst ^{2} c_{VVHH} V_\mu^aV^{\mu a} H^\dagger H
-\frac{g}2 c_{VVW}\epsilon_{abc}W^{\mu\nu a} V_\mu^b V_\nu^c\nn\\
\label{sml}
\eea
Here $V_\mu^a$, $a=1,2,3$, is the new heavy vector triplet, in the adjoint representation of $SU(2)_L$ and with vanishing hypercharge, describing the  charged and the neutral heavy spin-1 particles:
$V_\mu^\pm=(V_\mu^1\mp i V_\mu^{2})/{\sqrt{2}}$ and  $V_\mu^0=V_\mu^{3}$.
The covariant derivatives in \eq{sml} are defined by
\beq\label{Vcovder}
D_{[\mu}V^a_{\nu ]}=D_{\mu}V^a_{\nu } -D_{\nu}V^a_{\mu }\,,\;\;\;\;\; D_\mu V_\nu^a = \partial_\mu V_\nu^a
+g\,\epsilon^{abc}W_\mu^bV_\nu^c\,,
\eeq
where $g$ denotes the $SU(2)_L$ gauge coupling. The last two terms of the first row  of \eq{sml} contain the interactions of the new vector bosons with the Higgs and the three Goldstones (encoded in $H$) and therefore, by the Equivalence Theorem, with the longitudinal components of $W$ and $Z$ which are their relevant degrees of freedom at high energies, and with the SM fermions. Specifically:
\beq
i\,H^\dagger \tau^a {\overset{{}_{\leftrightarrow}}{D}}^\mu H=i\,H^\dagger \tau^a D^\mu H\,-\,i\,D^\mu H^\dagger \tau^a  H\,,
\eeq
\beq D^\mu H=(\partial^\mu +i g \f{\sigma^a}2 W^a-ig^\prime \f{\sigma^3}2 Y^\mu)H\,,
\eeq
with $g^\prime$  the weak hypercharge coupling,
while  $J_F^{\mu\;a}$ are the SM left-handed fermionic currents
\beq
\displaystyle
J_F^{\mu\;a}=\sum_f\overline{f}_L\gamma^\mu \f{\sigma^a}2f_L\,.
\eeq
The operators in the second line of \eq{sml} are not relevant for LHC because they do not contain vertices with SM gauge bosons. 

\begin{figure}[t!]
\centering
\resizebox{0.65\columnwidth}{!}{%
\includegraphics{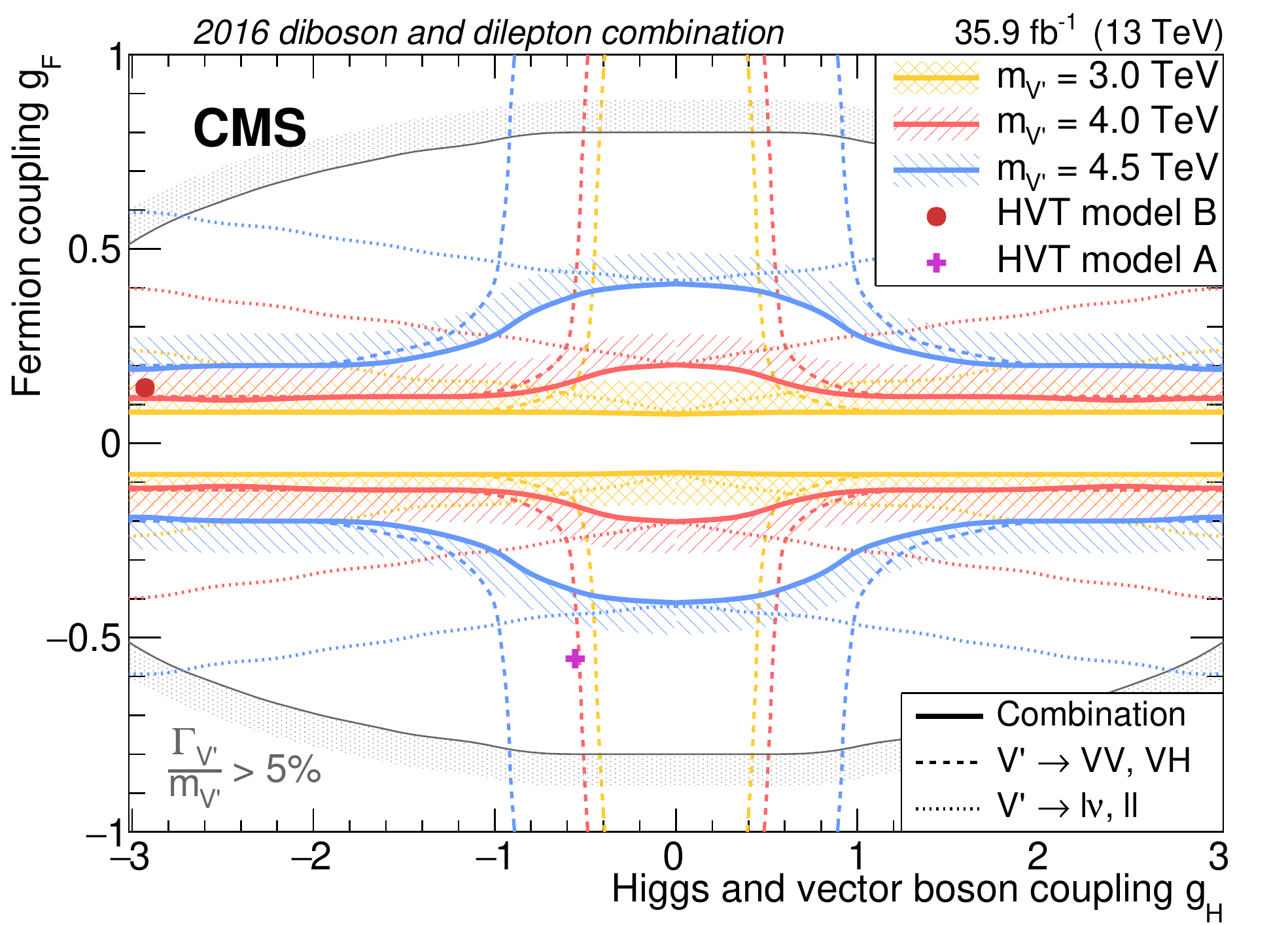} }
 \caption{Exclusion limits on the couplings of heavy vector resonances to fermions $g_F$ and SM vector bosons and the Higgs boson $g_H$ for narrow resonances obtained from the statistical combination of all the channels (solid lines). The dilepton (dotted lines) and the diboson searches (dashed lines)  constrain different regions. Three resonance masses hypotheses (3.0, 4.0, and 4.5 TeV) are considered. The hatched bands indicate the regions excluded. The analysis uses an integrated luminosity of 35.9 fb$^{-1}$. From  \cite{Sirunyan:2019vgt}. }
\label{fig:1}       
 \end{figure}

Analysis at ATLAS and CMS on new heavy vector triplet models are performed by using the first line operators, in terms of $m_V$ and two  parameters $g_H$ and $g_F$ defined as
\beq
g_H=g_Vc_H,\,\,\,\,g_F=\f {g^2 c_F}{g_V}\,.
\eeq
In Fig.~\ref{fig:1} the CMS bounds on $(g_H,g_F)$, obtained by using an integrated luminosity of 35.6 fb$^{-1}$
at the LHC with  a center of mass energy of 13 TeV \cite{Sirunyan:2019vgt} are shown. A  combination of searches for a heavy triplet of resonances decaying into pairs of vector bosons, a vector boson and a Higgs boson, two Higgs bosons, or pairs of leptons, has been considered. The bounds are obtained for narrow resonances, the areas bounded by the thin gray contour lines correspond to regions where the resonance widths are predicted to be larger than the average experimental resolution (5\%). The red dot and the violet cross correspond to  model A and B  that we review in the next subsection. 
A similar analysis has been performed by ATLAS \cite{Aaboud:2018bun}.
 
 \begin{figure}[b!]
\centering
\resizebox{0.6\columnwidth}{!}{%
\includegraphics{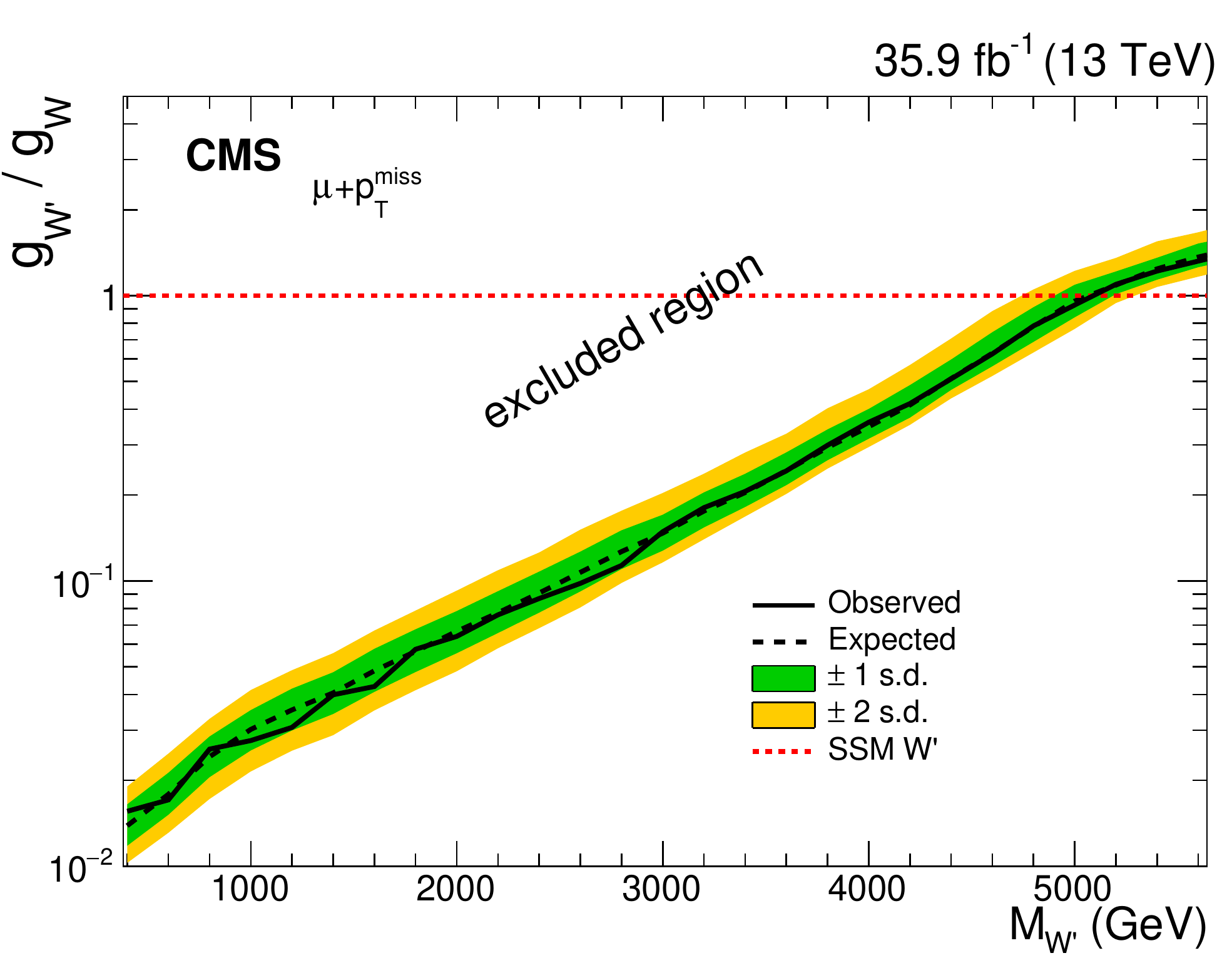} }
 \caption{Expected (dashed line) and observed (solid line) 95\%CL limits on the coupling strength  $g_{W^\prime}/g_W$ as  functions of the $W^\prime$ mass, for 
 the muon channel.
 The area above the limit line is excluded.  The SSM $W^\prime$ couplings are shown as a dotted line. The analysis uses an integrated luminosity of 35.9 fb$^{-1}$. 
 Similar bounds are extracted from the electron channel. 
 From \cite{Sirunyan:2018mpc}. }
\label{fig:2}       
 \end{figure}
In Fig.~\ref{fig:2}  we present a further result from a CMS analysis corresponding to the same integrated luminosity of 35.9 fb$^{-1}$, where only the charged channel in Drell-Yan process of new resonances has been considered to obtain 95\%CL limits on the coupling strength  $g_{W^\prime}/g_W$ as a function of the $W^\prime$ mass \cite{Sirunyan:2018mpc}.  Notice that the for the Sequential Standard Model (SSM), where the coupling of the $W'$ is set at the SM value, the  bound on $M_{W'}$ is around 5 TeV.

\section{Two reference models}
\label{sec:3}
In this section we review two reference models that have been proposed in the past to describe additional triplets and  have been used in the ATLAS and CMS analyses. They are a model, ({\bf A}),  with an extended electroweak symmetry \cite{Barger:1980ix}  and a strongly coupled composite model ({\bf B})\cite{Contino:2011np}.

\subsection{Model {\bf A}}
\label{sec:3_1}
This model corresponds to an extension of the electroweak symmetry to $SU(2)_1\otimes SU(2)_2\otimes U(1)_Y$ broken to $SU(2)\otimes U(1)_Y$. Fermions are supposed to transform only under  $SU(2)_1\otimes U(1)_Y$. Two scalar fields are introduced: the SM Higgs $H$, a doublet of $SU(2)_1$, and an additional one, $\Phi$, belonging to the representation $\bf{(2,2)}$ of $SU(2)_1\otimes SU(2)_2\otimes U(1)_Y$, the latter acquiring a vacuum expectation value:
\beq
\langle\Phi\rangle=\begin{pmatrix}f&0\\0&f
\end{pmatrix}.
\eeq
This bidoublet breaks the  $SU(2)_1\otimes SU(2)_2$ symmetry
to the diagonal $SU(2)$. The gauge coupling constants are identified as $g_V=g_2$, $1/g^2=1/g_1^2+1/g_2^2$, with $g_{1,2}$ the gauge couplings of $SU(2)_{1,2}$.
Other details  can be found in \cite{Barger:1980ix} and \cite{Pappadopulo:2014qza}. In Fig.~\ref{fig:1} the violet cross represents this model. Only masses of the new resonances higher than 4.5-5 TeV are allowed.

\subsection{Model {\bf B}}
\label{sec:3_2}
This  is the simplest model where
the Higgs boson is  realized as a pseudo Nambu-Goldstone from the breaking, at the scale $f$,  of an extended symmetry, $SO(5)\to SO(4)$, the so-called   Minimal Composite Higgs Model \cite{Contino:2011np}. The construction of the Lagrangian is based on the
Callan-Coleman-Wess-Zumino formalism. The Higgs is identified as a (2,2) under the unbroken  $SU(2)_L\otimes SU(2)_R\sim SO(4)$ and new vectors are introduced as representations  of the unbroken $SO(4)$. In particular a new triplet field $\rho_\mu$, $\bf{(3,1)}$ of $SU(2)_L\otimes SU(2)_R\sim SO(4)$ is present in the spectrum. 
 The Lagrangian describing the Higgs gauge boson sector is given by
\begin{equation}\label{rholag}
\mathcal L_{\rho}=-\frac{1}{4\tilde g'^2}(B_{\mu\nu})^2-\frac{1}{4 \tilde g^2}(W^a_{\mu\nu})^2+\frac{f^2}{4} d_\mu^{i}d^{\mu i}-\frac{1}{4 g_\rho^2}({\rho}_{\mu\nu}^{a})^2+\frac{m_{\rho}^2}{2g_\rho^2}\left({\rho}_\mu^{a}-e_\mu^{a}\right)^2.
\end{equation}
The $\rho$ field strength is given by $\dst {\rho}_{\mu\nu}^{a}= \partial_\mu {\rho}_{\nu}^{a}- \partial_\nu {\rho}_{\mu}^{a}-\epsilon^{abc}{\rho}_{\mu}^{b}{\rho}_{\nu}^{c}$, while the $d$ and ($\rho -e$) terms for the $SO(5)/SO(4)$ coset, in the large $f$ limit, are given by (see also Ref.~\cite{DeSimone:2012fs}):
\begin{equation}\label{dmudmu}
d_\mu^i d^{\mu\, i}= \frac{4}{f^2}|D_\mu H|^2+\frac{2}{3 f^{4}}\left[(\partial_\mu |H|^2)^2-4 |H|^2 |D_\mu H|^2\right]+O(\frac{1}{f^6}) \, ,
\end{equation}
and
\begin{equation}\label{emu}
\rho^a_\mu- e_\mu^a=\rho_{\mu}^{a}+W_\mu^a-\frac{i}{f^{2}}H^\dagger \tau^a \overset{\leftrightarrow}{D_{\mu}} H+\frac{i}{f^{4}}|H|^2H^\dagger \tau^a \overset{\leftrightarrow}{D_{\mu}} H+O(\frac{1}{f^6}).
\end{equation}
The heavy vector triplet $V_\mu^a$ of \eq{sml}, is identified as
\begin{equation}
V_\mu^a\equiv\rho_{\mu}^{a}+W_\mu^a.
\end{equation}

After the breaking of $SU(2)\otimes U(1)_Y$, the couplings of the physical Higgs to pairs of SM gauge bosons are rescaled by
\beq
\sqrt{1-\xi},\,\,\,{\rm with}\,\,\,\, \,\xi=\frac {v^2} {f^2},\,\,\,\,\,v=245~{\rm GeV}.
\eeq
The current limit on $\xi$, using $k_V=1.05\pm 0.04$ from ATLAS \cite{Aad:2019mbh}, is  $\xi \lessapprox 0.06$ at 95\%CL.

The coupling $g_V$ is here identified with $g_\rho$, and it turns out \cite{Pappadopulo:2014qza}
that 
\beq
c_H\sim c_F\sim 1\,.
\eeq
The red dot of  Fig.~\ref{fig:1} corresponds to  $c_H=-1$ and $g_V=3$, for which resonances ligher than  4-4.5 TeV are excluded. 
Other vector resonances like $\rho_R$ belonging to the representation $\bf{(1,3)}$ of $SU(2)_L\otimes SU(2)_R$ could be included in the model, see for instance \cite{Marzocca:2012zn}.

\section{Effective Lagrangians for new spin-1 triplets }
\label{sec:4}
A general procedure for building effective Lagrangians describing new vector resonances is the so-called {\it hidden local symmetry approach} \cite{Bando:1984ej,Casalbuoni:1985kq,Casalbuoni:1988xm}. We review here the main ingredients for two new triplets of vector resonances. For similar proposals in the framework of walking technicolor models see \cite{Appelquist:1999dq,Foadi:2007ue}. In these approaches the Higgs is described as an isosinglet scalar state.

We start considering the following group structure: $G'=SU(2)_L\otimes SU(2)_R \otimes SU(2)_L^\prime \otimes SU(2)_R^\prime$ broken to the diagonal $SU(2)_V$. The nine Goldstone bosons resulting from the spontaneous breaking 
 can be described by three independent
$SU(2)$ fields  $L(x)$, $R(x)$ and $M(x)$, transforming under the extended symmetry  group $G'$  as follows:
\beq
L'= g_L L h_L,~~~~~~R'= g_R R h_R,~~~~~~M'= h_R^\dagger M h_L\,,
\label{transf}
\eeq
with $g_{L,R}\in SU(2)_{L,R}$ and $h_{L,R}\in SU(2)_{L,R}^\prime$. 

In order to derive the most general Lagrangian up to second order in the derivatives, it is convenient to consider the following covariant quantities
\beq
{\cal V}_0^\mu=L^\dagger D^\mu L,\,\,\,
{\cal V}_1^\mu=M^\dagger D^\mu M,\,\,\,
{\cal V}_2^\mu=M^\dagger(R^\dagger D^\mu R)M\,,
\label{invts}
\eeq
with
\beq
\label{DD}
D_\mu L=\partial_\mu L -L \lmu ,\,\,\,\,
D_\mu R=\partial_\mu R -R \rmu ,\,\,\,
D_\mu M=\partial_\mu M -M \lmu+\rmu M
\eeq
where $\lmu=ig_L L_\mu ^a {\tau^a}/{2}$,
$\rmu=ig_R R_\mu ^a{\tau^a}/{2}$ are gauge fields of $SU(2)_L^\prime \otimes SU(2)_R^\prime$.
Using \eq{invts}
we can build six independent  invariants
under the transformations of \eq{transf}:
\beq
\label{inva}
\tr {\cal V}_0^2,\,\,\,\,\tr {\cal V}_1^2,\,\,\,\tr {\cal V}_2^2,\,\,\,
\tr ({\cal V}_0 {\cal V}_1),\,\,\,\tr ({\cal V}_1{\cal V}_2),\,\,\,
\tr ({\cal V}_0 {\cal V}_2)\,.
\eeq
In conclusion, the most general Lagrangian for the vector boson interactions is a combination of these six invariants. The gauging of the Lagrangian with respect to the electroweak symmetry is obtained by substituting the covariant derivatives of \eq{DD} with
\beq
D_\mu L \to D_\mu L+ {\Wt}_\mu L,\,\,\,
D_\mu R \to D_\mu R+ {\Yt}_\mu R,\,\,\,\,
D_\mu M \to D_\mu M \,,
\label{cov}
\eeq
where
$
\Wt_\mu=i\gt {\tilde W}_\mu ^a{\tau^a}/{2}$,
$\Yt_\mu=i\gpt
 {\tilde Y}_\mu {\tau^3}/{2}$, and the "tildes"  indicate quantities which are not the SM ones due to mixing.
The  non-linear formulation of the SM is obtained in terms of the matrix $U\equiv LM^\dagger R^\dagger$ and the scalar singlet $h$.
The effective Higgs  Lagrangian is then written in terms of the quantities given in \eq{invts} as
\bea
\label{HV}
\frac{v^2}{4}(1+2a\f h v  +b \f {h^2}{v^2} )&&\tr (D^\mu U)^\dagger D^\mu U+...\nn\\ =
&&-\frac{v^2}{4} (1+2 a\f h v  +b \f {h^2}{v^2} )\tr[({\cal V}_0-{\cal V}_1-{\cal V}_2)^2]+...
\eea
The SM couplings correspond to $a=b=1$ and   dots stand for  interaction terms between  the scalar singlet $h$ and the other invariants of \eq{inva}.
For example, by adding  the two invariants $\tr{\cal V}_0^2$ and $\tr{\cal V}_2^2$,  one obtains a model with two spin-1 vector triplets, with $\lmu$ and $\rmu$ unmixed, when neglecting weak interactions.

Following the {\it hidden local symmetry} approach, here briefly reviewed, the SM fermions are coupled to the extra spin-1 resonances via mixing with the SM gauge bosons. In addition, direct couplings, for example, of the $\psi_L$ fermions  to $\lmu$  are allowed \cite{Casalbuoni:1985kq}.  In fact, for each, $\psi_L$ we can
construct a $SU(2)$ doublet
 \beq \chi_L=L^\dagger
\psi_L\,,\,\,
\label{eq:8} \eeq
and consider the following invariant term 
\beq  b_L
\bar\chi_L i\gamma^\mu (\de_\mu+i \lmu+ \frac i 2\gpt (B-L)
Y_\mu)\chi_L\eeq where $b_L$ is a  dimensionless parameter. 
In
the unitary gauge $L=I$ and, shifting $\psi_L\rightarrow \psi_L/{\sqrt{ 1+b_L}}$,
one gets a fermion-heavy triplet interaction  term
\beq
 -\frac {b_L}{{1+ b_L}}\bar\psi_L \gamma^\mu  \lmu \psi_L
\eeq
contributing, together with the mixing term, to $g_F=g^2 c_F/g_V$ in \eq{sml}. 
The procedure can be generalized to several triplets \cite{Casalbuoni:2005rs}.

\subsection{Explicit models with two new spin-1 triplets}
\label{foursite}

In the following we will concentrate on two different models describing two extra triplets of spin-1 resonances. We have seen that their interactions are described, in general, by six independent invariants.
They are reduced to four by requiring  the
invariance under the discrete left-right transformation, denoted by $P: L\leftrightarrow R,~M\leftrightarrow M^\dagger
$.  Namely, the most general $G'\otimes P$ invariant Lagrangian is given by \cite{Casalbuoni:1988xm}
\beq
{\cal L}_G=-\frac{v^2}{4} [a_1 I_1 + a_2 I_2 + a_3 I_3 + a_4 I_4]+ {\cal L}_{kin}
\label{lg}
\eeq
with
\bea
I_1&=&\tr[({\cal V}_0-{\cal V}_1-{\cal V}_2)^2],\,\,\,
I_2=\tr[({\cal V}_0+{\cal V}_2)^2],\nn\\
I_3&=&\tr[({\cal V}_0-{\cal V}_2)^2],\,\,\,\,
I_4=\tr[{\cal V}_1^2]
\eea
and the kinetic terms (we take $g_L=g_R={\gs}/{\sqrt{2}}$):
\beq
{\cal L}_{kin}=\frac{1}{\gs^2} \tr[F_{\mu\nu}({\bf L})]^2+
	 \frac{1}{\gs^2}  \tr[F_{\mu\nu}({\bf R})]^2\,.
\eeq

\subsubsection{4-Site Model with a Composite Higgs}

As said, new vector resonances appear in five-dimensional extensions of the SM as Kaluza-Klein (KK) excitations of the SM gauge bosons \cite{Agashe:2003zs,Csaki:2003dt,Csaki:2003zu}. When deconstructed  \cite{ArkaniHamed:2001ca,ArkaniHamed:2001nc,Hill:2000mu,Cheng:2001vd}, these theories appear as gauge theories with extended $SU(2)$ symmetries.

We review here a simple four-dimensional model  \cite{Accomando:2008jh,Accomando:2010ir,Accomando:2011eu,Accomando:2011xi,Accomando:2012yg,Fedeli:2012cn}, corresponding to a deconstruction including two additional copies of the $SU(2)$ symmetry, namely $G'=SU(2)_L\otimes SU(2)_R \otimes SU(2)_1 \otimes SU(2)_2$ with the particular choice:
\beq
a_1=0,\,\,\,a_2=a_3=\frac {2 f_1^2}{v^2},\,\,\,\,a_4=\frac {4 f_2^2}{v^2},\,\,\,\,
\frac 4{ v^2}=\frac 2 {f_1^2}+\frac 1 {f_2^2}\,.
\label{v}
\eeq
Following the general construction illustrated above, we add to the Lagrangian the  scalar-scalar and scalar-vector interactions as in \eq{HV} (with scalar we here refer both the Higgs particle and the Goldstones):
\bea
{\cal L}_{hG}&=&
 (2 a_h \frac h v +b_h \frac {h^2}{v^2})
f_1^2[\tr (D_\mu \Sigma_1)^\dagger D^\mu \Sigma_1+\tr (D_\mu \Sigma_3)^\dagger D^\mu \Sigma_3]\nn\\
&+& (2 c_h \frac h v +d_h\frac {h^2}{v^2}) f_2^2\tr (D_\mu \Sigma_2)^\dagger D^\mu \Sigma_2\,,
\label{LhG}
\eea
with the  identification of the chiral fields:
$
\Sigma_1=L,\,\,\Sigma_2=M^\dagger,\,\,\Sigma_3=R
$. The covariant derivatives are defined as in \eq{cov} with the additional gauge fields identified as: $\lmu\to -{\bf \tilde{A}}_\mu^{1}$, $\rmu\to -{\bf \tilde{A}}_\mu^{2}$,
${\bf\tilde{A}}_\mu^i=ig_i\tilde{A}_\mu^{ia}\tau^a/2$   and we assume $g_1=g_2$ for the couplings.

In the unitary gauge, this model predicts two new triplets of gauge 
bosons, which acquire mass through the same non-linear symmetry breaking 
mechanism giving mass to the SM gauge bosons. For large $g_1$,  $W^\pm,Z$ mass eigenvalues are given by
\beq
M^2_{W^\pm}\simeq\frac {\gt^2 v^2}4 \left [1-\frac {\gt^2}{2g_1^2}(1+z^4)\right ],\,\,\,\,M^2_{Z}\simeq \frac {(\gt^2+{\t g}^{\prime 2})v^2}4 \left [1-\frac {\gt^2}{2g_1^2}z_\zeta\right ]\,,
\eeq
where
\beq
z=\frac {f_1} {\sqrt{f_1^2+2 f_2^2}},\,\,\, z_\zeta=\frac 1 2 \frac {z^4+\cos^2 2\t \theta}{\cos^2\t \theta},\,\,\,\tan\t\theta=\frac \gpt \gt\,.
\eeq
The heavy vector triplets, ${\bf A}^1=(W_1^\pm, Z_1)$ and  ${\bf A}^2=(W_2^\pm, Z_2)$, have masses approximately given by $M_{1}, M_{2}$ with
\beq
M_1\simeq \frac {g_1v}{\sqrt{2(1-z^2)}},\,\,\,\,M_2\simeq \frac 1 z M_1\,,
\eeq

As said, direct couplings  of the extra spin-1 resonances to the SM fermions are allowed by the symmetries. They are present also in the small mixing limit and can be introduced by following the construction described above.

From \eq{LhG} we derive the 
 couplings of the Higgs $h$  to the charged
gauge bosons:
\beq
\frac {2h}{v}  \left [a M^2_W W^+ W^- + a_h M_1^2 W^+_1W^-_1
+ (a_h z^2+c_h (1-z^2)M_2^2 W^+_2W^-_2\right ]\,,
\eeq
\beq
a=a_h(1+z^2)+c_hz^2\,,
\eeq
and similarly for the neutral sector.
The SM limit is obtained for $z\to 1$ (namely $g_1,f_1\to\infty$), $c_h\to 1$ and $a_h\to 0$. In this limit $a\to 1$.

\begin{figure}[h!]
\centering
\resizebox{0.85\columnwidth}{!}{%
\includegraphics{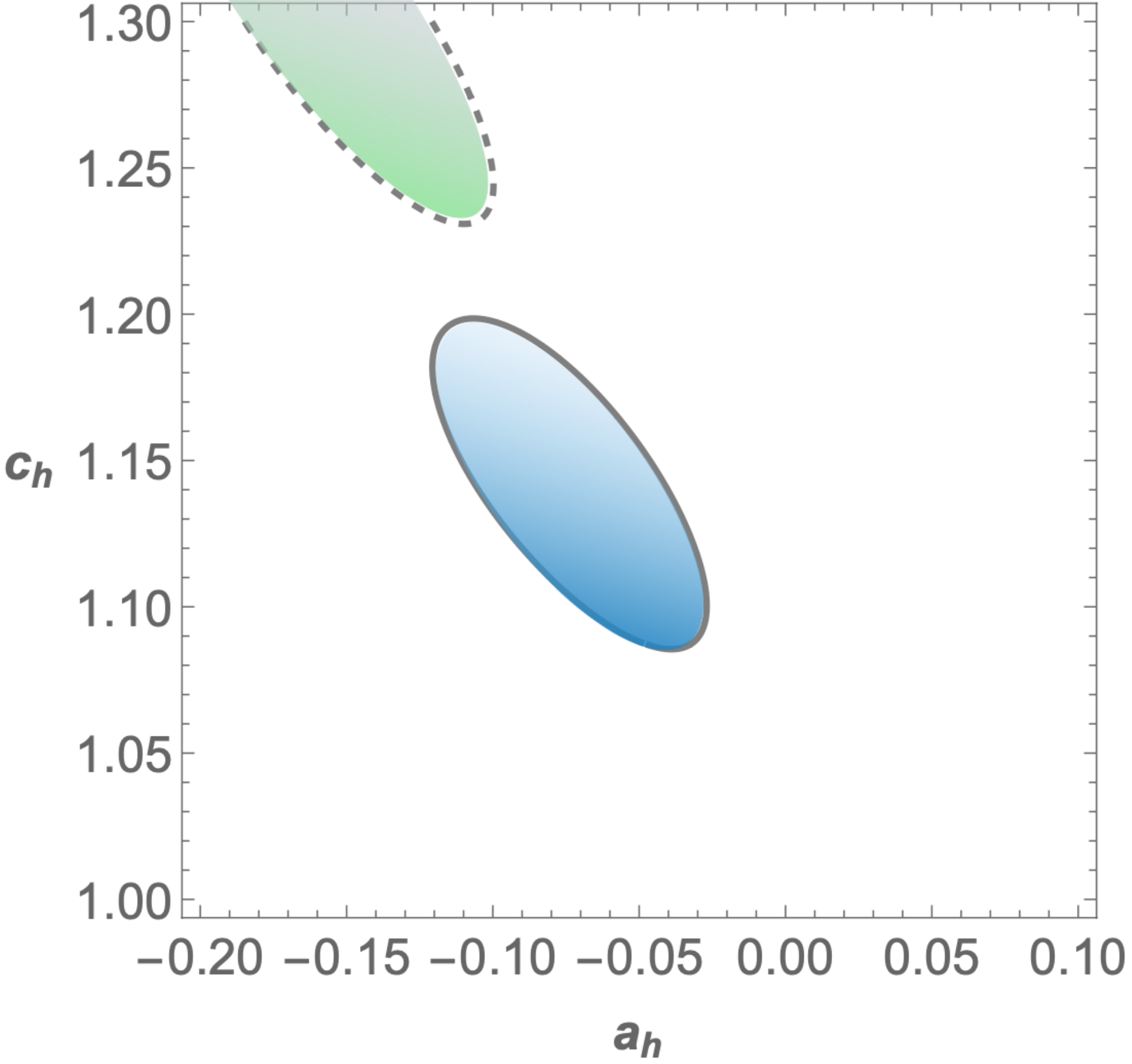}\hspace{1.2cm}
\includegraphics{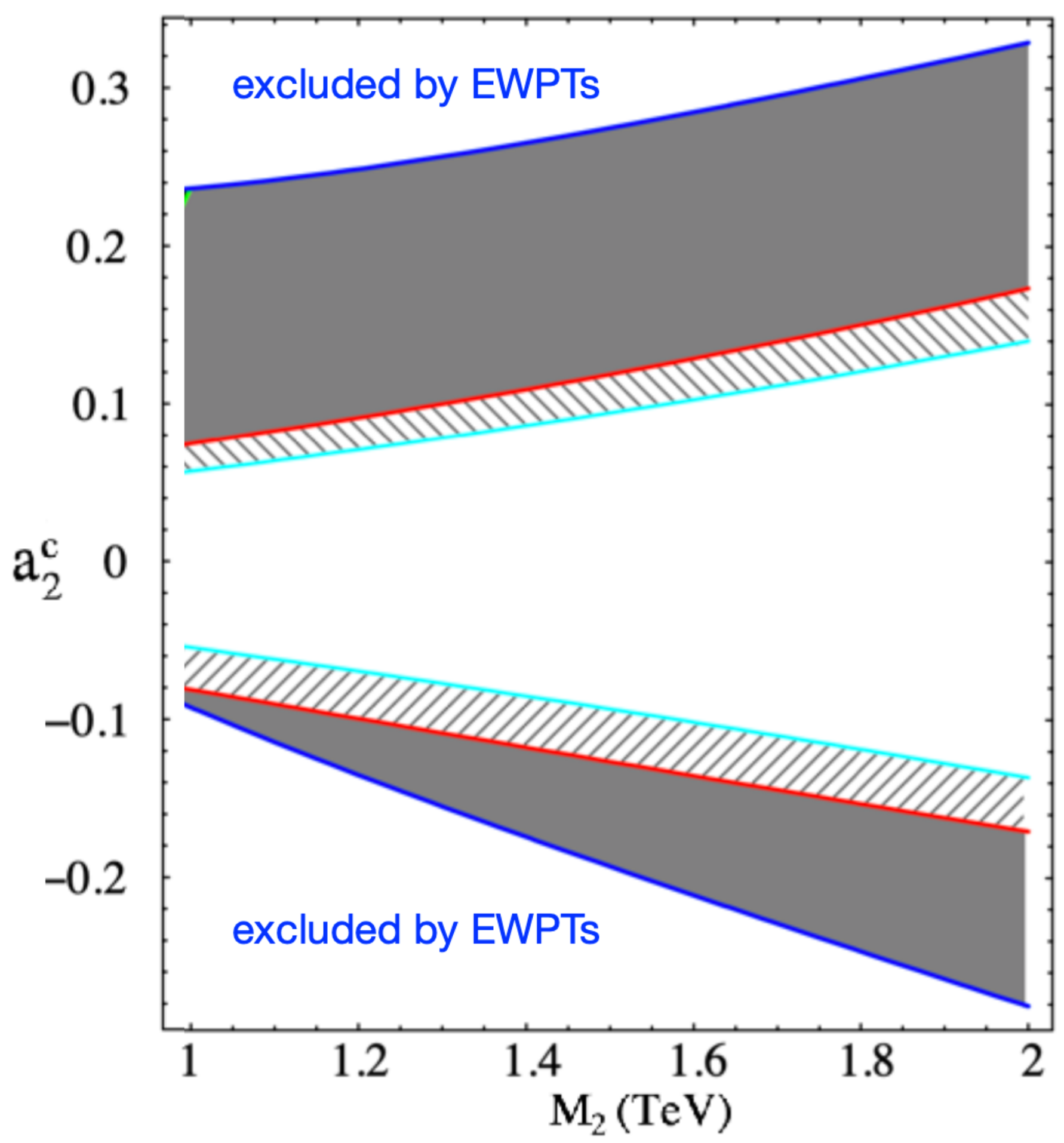} }
 \caption{{\it Left:} 95\%CL bounds on the plane $(a_h,c_h)$ for $z=0.95$ (continuous blue) and $z=0.9$ (dashed green)
 from ATLAS measurements \cite{Aad:2019mbh} on $k_V=1.05\pm 0.04$, $k_\gamma=1\pm 0.06$ based on 79.8~fb$^{-1}$. {\it Right:} 5$\sigma$-discovery plot from D-Y production at LHC with L=100 fb$^{-1}$ for $z=0.8$ in the plane
 $(a^c_{2}, M_2)$.  Inside the grey (dashed) (central white) regions both $W_{1,2}$ (only $W_2$) (no resonance) can be detected.  From \cite{Accomando:2008jh}.}
 \label{fig:3}       
 \end{figure}

Bounds on $hW^+W^-$, $hZZ$, $h\gamma\gamma$ couplings from recent analysis at LHC, can be translated in limits on $a_h,c_h$ for different values of $z$.  Assuming no deviation in the top Higgs couplings ($c_t= 1$), we can compute \cite{Bellazzini:2012tv}
\beq
\Gamma/\Gamma_{SM}(h\to\gamma\gamma)\sim \left [1+ \frac 9 8 [a_h (1-z^2)+c_h(1-z^2)] +\frac 9 7 (a-1)\right ]
\eeq
and use ATLAS measurements \cite{Aad:2019mbh} on $k_V=1.05\pm 0.04$, $k_\gamma=1\pm 0.06$ based on 79.8~fb$^{-1}$ of integrated luminosity to get the 95\%CL bounds shown in Fig.~\ref{fig:3} (Left).

Unitarity bounds on the 4-Site model extra spin-1 resonance masses are approximately  2-4 TeV depending on the $z$ value. For $z=0.95$ unitarity bounds increase to approximately 4 TeV \cite{Accomando:2012yg}.

Bounds from $S,T,U$ parameters on the charged couplings of the heavy triplets to SM fermions, $a_{1}^c $ and $a_{2}^c$, and on the corresponding neutral couplings,  can be found in \cite{Accomando:2008jh,Accomando:2010ir,Accomando:2011eu,Accomando:2011xi,Accomando:2012yg,Fedeli:2012cn}. For example, for $z=0.95$, $M_{W_1}=2$ TeV, one obtains $-0.35\leq a_{1}^c \leq 0.4$.
An extended analysis of Drell-Yan production of these new vectors in the dilepton and diboson channel at LHC was also performed in \cite{Accomando:2008jh,Accomando:2010ir,Accomando:2011eu,Accomando:2011xi,Accomando:2012yg,Fedeli:2012cn}.  The 5$\sigma$-discovery contours for a foreseen LHC integrated luminosity of 100 fb$^{-1}$  in the plane $(a^c_2,M_2)$, for $z=0.8$, where $a_2^c$ is the $W_2$
charged coupling and $M_2$  is approximately $W_2$ mass, are plotted in  Fig.~\ref{fig:3} (Right). The upper and lower parts of the plot are excluded by the EWPTs. Inside the grey regions both
$W_{1,2}$ are visible; inside the dashed ones only $W_2$ can be
detected. Inside the central uncolored region no resonance is
visible in the Drell-Yan channel.
 
However, in presence of several new triplets, as in this model, we cannot extract bounds on the couplings of the new resonances from Fig.~\ref{fig:1} and Fig.~\ref{fig:2}  because as shown in \cite{Accomando:2011eu,Accomando:2013sfa,Accomando:2016mvz} there are significant interference effects that cannot be neglected. In Section \ref{sec:5} we review these effects in a composite Higgs model.


\subsubsection{Linear degenerate BESS}
A second interesting choice in the general parameterization  given in \eq{lg} is $a_1=1$, $a_4=0$, $a_2=a_3$.
The Lagrangian then reads
\beq
{\cal L}_G=\frac{v^2}{4}\{ \tr(\partial_\mu U^\dagger \partial^\mu U) + 2~a_2~ [\tr(D_\mu L^\dagger D^\mu L)+ \tr(D_\mu R^\dagger D^\mu R)]\}\,.
\label{a2}
\eeq
In this model, before turning on electro-weak interactions, the new triplets $\lmu$ and $\rmu$ are degenerate in mass (from here the name {\it degenerate Breaking the Electro-Weak Symmetry Strongly (BESS) Model}).
Notice that each of the three terms in the above expression
is invariant under an independent $SU(2)\otimes SU(2)$
group
\beq
U'=\omega_L U \omega_R^\dagger,~~~~~~L'= g_L L h_L,~~~~~~R'= g_R R h_R\,,
\eeq
so that the symmetry of the Lagrangian is enlarged to $G_{max}=[SU(2)\otimes SU(2)]^3$. As a consequence of this enlarged symmetry or equivalently of the degeneracy of $\lmu$ and $\rmu$ fields  (or the combination corresponding to the vector and axial-vector fields) the correction to the $S$ parameter at leading order is zero \cite{Casalbuoni:1995qt}. The possibility of an enhanced symmetry in near conformal theories and the emergence of parity doublet was suggested also in \cite{Appelquist:1999dq} (see also \cite{Foadi:2007ue}) in the framework of walking technicolor. 

The degenerate BESS model has also a linear (renormalizable) formulation \cite{Casalbuoni:1996wa,Casalbuoni:1997rs}. It  is based on a gauge group 
$G=SU(2)_L\otimes U(1)\otimes SU(2)_{L}^\prime \otimes SU(2)_{R}^\prime$
and has a scalar sector consisting of scalar fields $\tilde L\in (2,0,2,0),
\tilde U\in (2,2,0,0),
\tilde R\in (0,2,0,2)$ of the group $G$ (it is a generalized version of Model {\bf A} in Sect.~\ref{sec:3_1}).
The scalar fields break the gauge
symmetries through the following chain 
\beq
\begin{matrix}
SU(2)_L\otimes U(1)\otimes SU(2)_{L}^\prime \otimes SU(2)_{R}^\prime\cr
\downarrow u\cr
SU(2)_{\rm weak}\otimes U(1)_Y\cr
\downarrow v\cr
U(1)_{\rm em}
\end{matrix}
\eeq

The two breakings are induced by the expectation values 
$\langle\tilde L\rangle=
\langle\tilde R\rangle=u$,  and $\langle\tilde U\rangle=v$ respectively and we 
assume  $u\gg v$.
The first two expectation 
values make the breaking $SU(2)_L\otimes SU(2)'_L\to SU(2)_{\rm weak}$ and  
$U(1)\otimes 
SU(2)'_R\to U(1)_Y$, whereas the second breaks in the standard way
$SU(2)_{\rm weak}\otimes U(1)_Y\to U(1)_{\rm em}$. 
The parameter $a_2$ of \eq{a2} is related to $u$ via $a_2=u^2/(2 v^2)$.

Standard fermions are supposed to couple only to $\tilde  U$.  The model, in the limit of large $u$, satisfies decoupling giving back the SM with a light Higgs and suppressed contributions to the parameters $S,T,U$.

We parameterize the scalar
fields as $\tilde L=\tilde\rho_L L$, $\tilde R= \tilde\rho_R  R$,
$\tilde U=\tilde\rho_U  U$, with ${L}^\dagger {L}=I$,
${R}^\dagger{R}=I$
 and ${U}^\dagger  {U}=I$.
The Lagrangian for their kinetic terms and interactions with the gauge bosons 
is given by ($\lmu\to {\bf \tilde{A}}_\mu^{1}$, $\rmu\to -{\bf \tilde{A}}_\mu^{2}$)
\beq
{\cal L}_{\rho G} = \frac 1 4 \left [ \tr (D_\mu \tilde U)^\dagger  (D^\mu\tilde U)
+\tr (D_\mu \tilde L)^\dagger  (D^\mu \tilde L)
+\tr (D_\mu \tilde R)^\dagger  (D^\mu \tilde R) \right ]      \,,     
\eeq
\bea
& &D \tilde L =\partial \tilde L +
\Wt_\mu\tilde L
-\tilde L {\bf\tilde{A}}_\mu^1\nn\\
& &D \tilde R =\partial\tilde R +\Yt_\mu\tilde R
-\tilde R{\bf\tilde{A}}_\mu^2\nn\\
& &D\tilde U =\partial\tilde U +
\Wt_\mu \tilde U   -\tilde U\Yt_\mu
\eea

The scalar potential is  expressed in terms of three Higgs
fields: 
 \beq \label{potential} 
 V= 2 \mu^2
(\rhot_L^2+\rhot_R^2)+ \lambda (\rhot_L^4+\rhot_R^4) +2 m^2
\rhot_U^2 +
 h \rhot_U^4
+ 2 f_3 \rhot_L^2\rhot_R^2 + 2 f \rhot_U^2(\rhot_L^2+\rhot_R^2)
\eeq

After shifting the fields by their VEV's $u,v\neq 0$, the squared mass
eigenvalues of the Higgses are: \bea\label{masses}
 m_{\rho_U}^2
\simeq 8\left (h-2\frac {f^2}{\lambda + f_3} \right )v^2,~~~
 m_{\rho_L}^2 \simeq 8  {(\lambda - f_3)}{u^2},~~~
m_{\rho_R}^2 \simeq 8( {\lambda + f_3}){u^2}.
\eea
Concerning the gauge sector, let's call the mass eigenstates of the new spin-1 resonances ${\bf A}^i=(W_i^\pm,Z_i)$, $i=1,2$.

The fields ${W}_2^{\pm}$ are unmixed and their squared mass is given by
\beq
M^2_{W_2}=\frac 1 4 g_1^2 u^2\equiv M^2
\eeq
where $g_1$ is the common  coupling constant of the extra gauge symmetry.
The other two charged eigenvalues, in the limit of small  $r= g^2 v^2 /(4 M^2)$, are 
\beq
M^2_{{W}}\simeq\frac \vv 4 g^2 (1-r \sf^2 ),\,\,\,\, M^2_{W_1}\simeq
 \frac {M^2} {\cf^2}\,,
\label{mw2}
\eeq
with $\varphi$  defined by the relation  $g = g_1 s_{\varphi}$, with $1/g^2=1/\gt^2+1/g_1^2$.
Notice that  the  SM  gauge boson masses
receive corrections $O(r)$  due to mixing.

In the charged sector,  at the first order in $r$, the couplings are given by
\beq
{\cal L}_{\rm fermions}^{\rm charged}
=-(a_W W_\mu^-+a_1^c W^{-}_{1\mu}) J_L^{\mu -}+h.c.,
\label{lch}
\eeq
with
\beq
a_W=\frac g {\sqrt {2}}(1-\sf^2 r),\,\,\,\,
a_1^c=-\frac g {\sqrt {2}}(1+\cf^2 r)\tan\varphi,
\eeq
and $J^\pm_L= \bar \psi_L \gamma^\mu \tau^\pm \psi_L$.
There is no coupling of $W_2^\pm$ to fermions, because
they do not mix with  $W^\pm$.  
The fermionic couplings of the neutral gauge boson sector are given in
\cite{Casalbuoni:1997rs}. The heavy gauge bosons are coupled to
fermions only through mixing with the SM ones, namely these couplings vanish for   $s_\varphi \to 0$. Nevertheless one could consider also direct couplings to fermions as shown in the derivation of the general effective Lagrangian.
\begin{figure}[t]
\centering
\vspace{-3.2cm}
\resizebox{0.7\columnwidth}{!}{%
\includegraphics{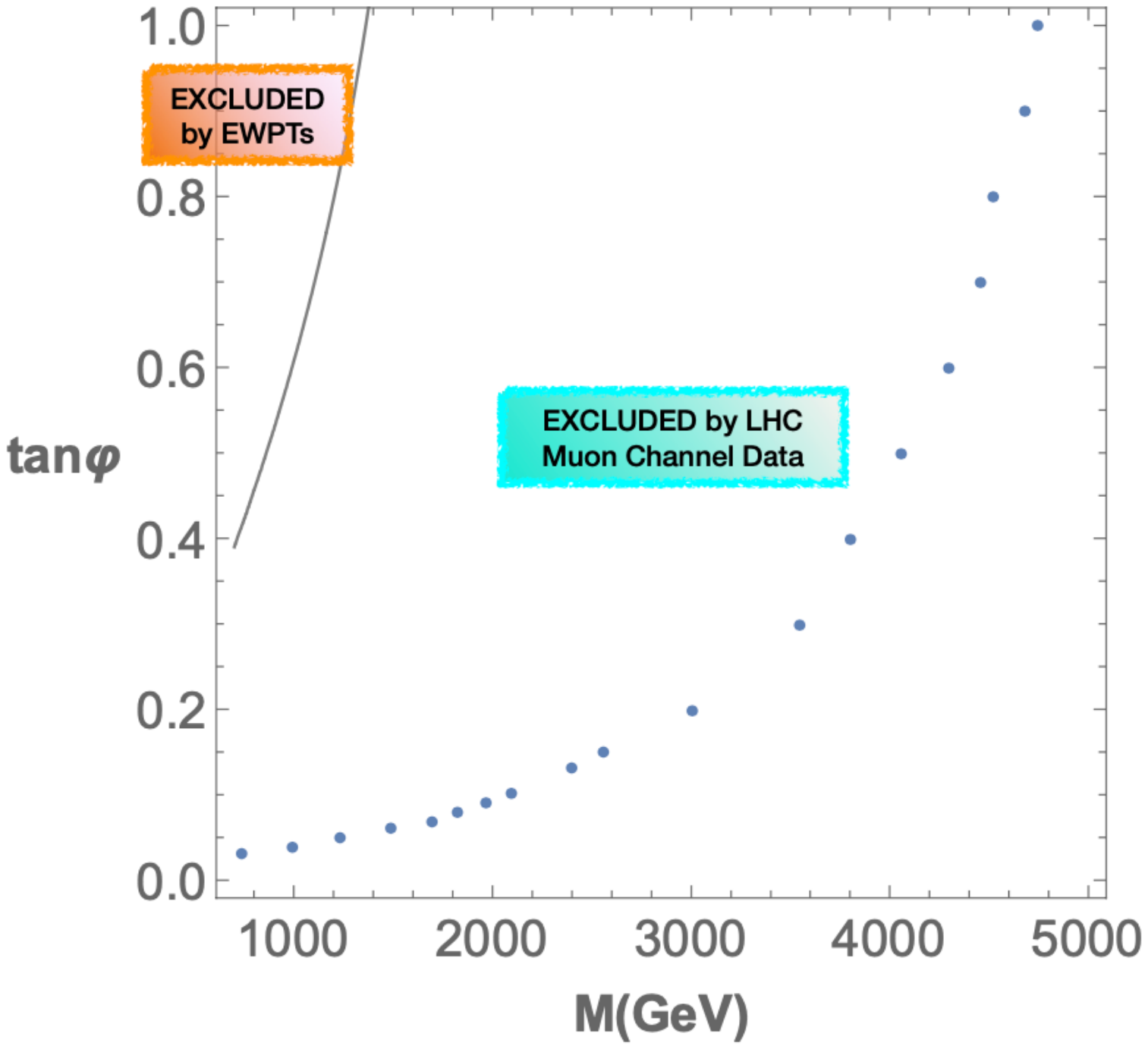} }
\vspace{-3.2cm}
 \caption{95\%CL bounds on the plane $(M,\tan\varphi)$ from electro-weak precision measurements (continuous line)  compared with  bounds (dotted) from the muon channel at LHC extracted from the analysis of Fig.~\ref{fig:2}. }
 \label{fig:4}       
 \end{figure}
 
 Since in this set-up, only the charged gauge bosons $W_1^\pm$ are coupled to fermions, there is no the interference effect between  $W_1^\pm$ and $W_2^\pm$  and we can use the analysis presented in Fig.~\ref{fig:2} for a single heavy triplet resonances to extract  bounds on the linear degenerate BESS  model. Results on the plane $(M,\tan\varphi)$ are shown in Fig.~\ref{fig:4}.

In the neutral sector both $Z_1$ and $Z_2$ are coupled to fermions 
and a dedicated analysis is necessary to get bounds from LHC results.
Couplings of the SM-like Higgs $\rho_U$ to SM fermions are rescaled by $c_\alpha$, where $\alpha$ is the mixing angle in the scalar sector, while the couplings of  $\rho_U$ to SM gauge bosons are given by
\beq
\rho_U\Big [\frac v 2 g^2 c_\alpha (1-s_\varphi^4x^2) W^+W^-+\frac v{4c_\theta^2}g^2c_\alpha
[1-2x^2\frac {s_\varphi^4}{c_\theta^4}(1-2c_\theta^2s_\theta^2)]Z^2\Big ]\,.
\eeq
 Due to the stringent limits on the plane $(M,\tan\varphi)$, current bounds from LHC on the parameter $k_V$ \cite{Aad:2019mbh} can be directly translated on a bound on $\alpha$, namely:  $0.97 \leq c_\alpha\leq 1$ at 2$\sigma$ level.

\section{Drell-Yan production and interference effects in models with multi-triplet structure}
\label{sec:5}

Drell-Yan production in the dilepton channels is a very good tool for searching new heavy triplets as we have already discussed in Section \ref{sec:2}. A class of BSM scenarios, like composite Higgs models and the SM model in five dimensions, predict several new spin-1 resonances. In these cases the interpretation of the experimental results and related bounds on the parameter space of the models is complicated by the presence of finite width and interference effects. We briefly review in this Section some  of such effects for the D-Y process of the neutral channels (new $Z^\prime$s) in composite Higgs models \cite{Accomando:2016mvz}. The four-dimensional composite Higgs model (4DCHM), that we are going to review,  is the one proposed in \cite{DeCurtis:2011yx}, based on two non-linear $\sigma$ models, one for $SO(5)/SO(4)$ and the second for $SO(5)_L\otimes SO(5)_R/SO(5)_{L+R}$. After the breaking  one is left with 4 neutral and 6 charged spin one heavy particles, or two triplets ({\bf 3,1})+({\bf 1,3}), almost degenerate in mass, and, in addition, two neutral and one charged coset resonances.

\begin{figure}[h]
\centering
\resizebox{1.\columnwidth}{!}{%
\includegraphics{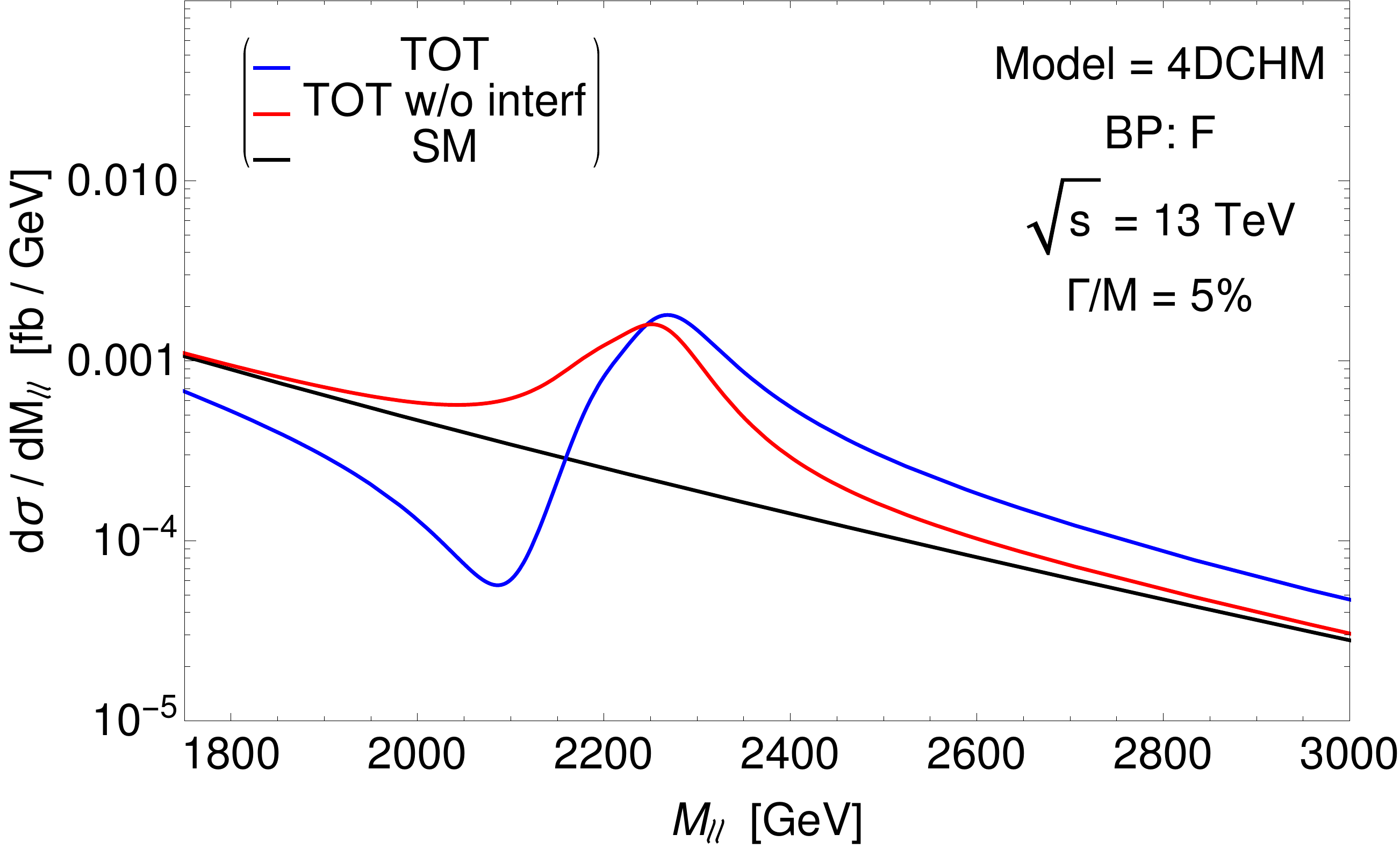}
\includegraphics{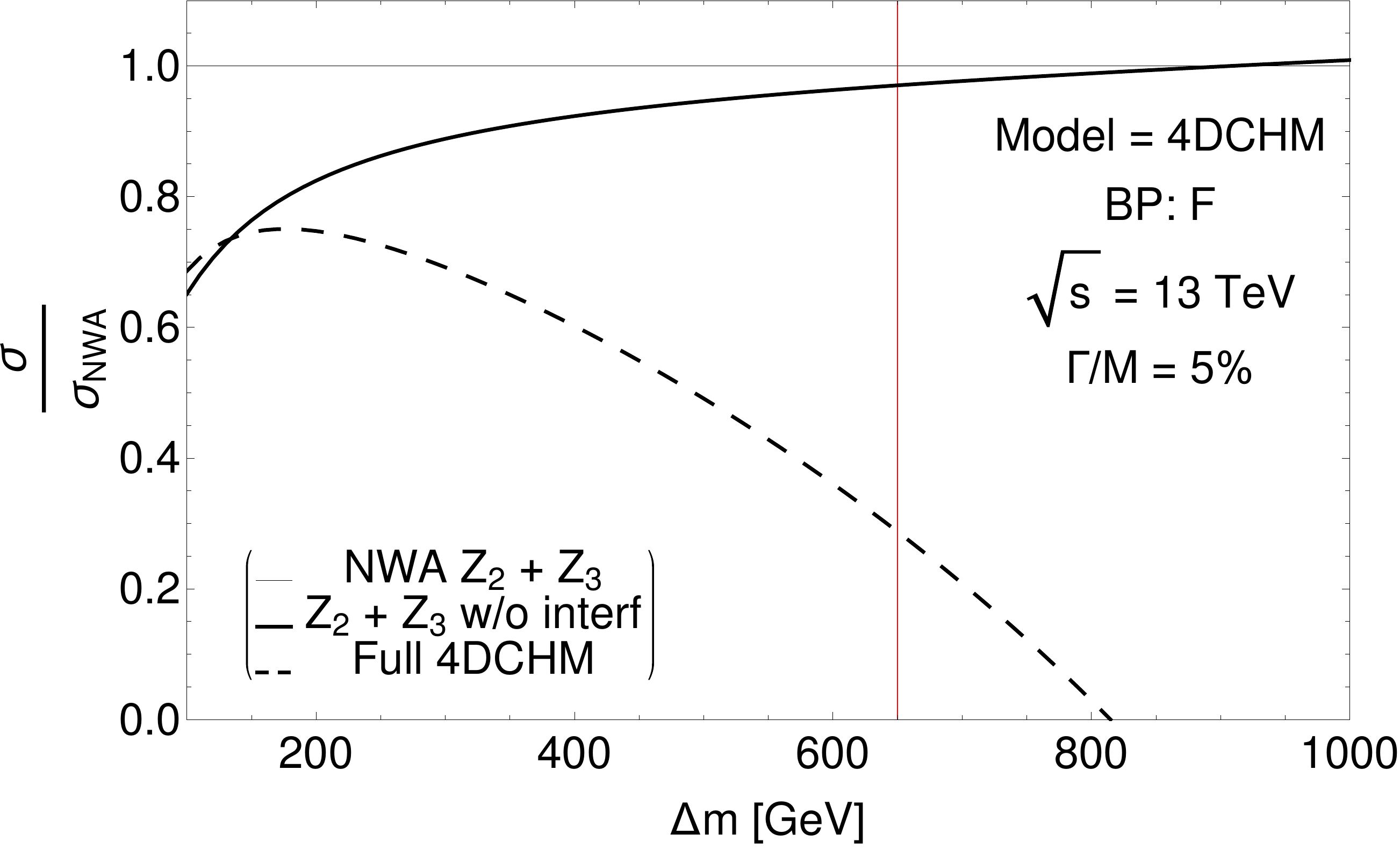}}
 \caption{(Left) Differential cross section in the dilepton invariant mass for the benchmark point F  within the  4DCHM i.e. double-resonant $Z_{2,3}$ scenario. (Right) Ratio of the full signal cross section for the $Z_{2,3}$ bosons corresponding to benchmark F within the  4DCHM scenario (dashed line) 
and the two resonances FWA (solid line) over the NWA result as a function of the symmetric integration interval around the peak. The vertical red line represents the CMS adopted optimal cut, $\delta m=|M_{ll}-M_{Z^\prime}|\leq 0.05 E_{LHC}$, which keeps the interference and FW effects below 10\% in the case of narrow single $Z^\prime$ models \cite{Accomando:2013sfa}. From \cite{Accomando:2016mvz}.}
\label{fig:5}       
 \end{figure}

The extra neutral resonances, that couple to fermions, are denoted by $Z_2$, $Z_3$, $Z_5$
 with squared masses, at order $O(\xi)$, given by
\begin{equation}
\begin{split}
 \begin{split}
& M^2_{Z_2}\simeq \frac{ m_\rho^2}{c_\psi^2} \left(1-\frac{s_\psi^2 c_\psi^4}{4 c_{2\psi}}\xi \right),\\
& M^2_{Z_3}\simeq \frac{ m_\rho^2}{c_\theta^2} \left (1-\frac{s_\theta^2 c_\theta^4}{4 c_{2\theta}}\xi\right)\\
& M^2_{Z_5}\simeq 2 m_\rho^2 \left[1+\frac 1 {16} (\frac 1 {c_{2\theta}}+\frac 1{2 c_{2 \psi}})\xi\right],
\end{split}
\end{split}
\label{eq:masses}
\end{equation}
with $\xi=v^2/f^2$, $\tan\theta=\tilde g/g_\rho$ and $\tan\psi=\sqrt{2} \tilde g'/g_{\rho}$ where $g_\rho$ is the gauge coupling of the $SO(5)$ group, $f$ is the scale of the spontaneous strong symmetry breaking and $m_\rho=f g_\rho$.  The $Z_5$ resonance, having a  mass $\sim \sqrt{2}$ higher  than $Z_2$ and $Z_3$, gives a negligible contribution to the Drell-Yan process at LHC RunII and it has been neglected in the analysis.
The change in the differential D-Y cross section in the dilepton invariant mass  produced by the interference is shown in Fig.~\ref{fig:5} for the benchmark F, corresponding to $M_{Z_2}=2192$ GeV and  $M_{Z_3}=2258$ GeV. The interference effects produce  the dip before the resonant peak, and spoil the analysis performed within the narrow width approximation (NWA)  or in the finite width approach (FWA). The same conclusion can be derived by plotting the ratio of the cross section over the NWA as a function of $\delta m=M_{ll}-M_{Z^\prime}$.  In the NWA the signal rate is typically estimated modelling the signal  as a Breit-Wigner function convoluted with a Gaussian function, which is used to describe the dilepton mass resolution. This example shows that one should avoid to use NWA or FWA within the 4DCHM or similar models like the 4-Site model of Section \ref{foursite} to extract  bounds on $Z^\prime$ couplings and  masses.

\begin{figure}[h]
\centering
\resizebox{1.01\columnwidth}{!}{%
\includegraphics{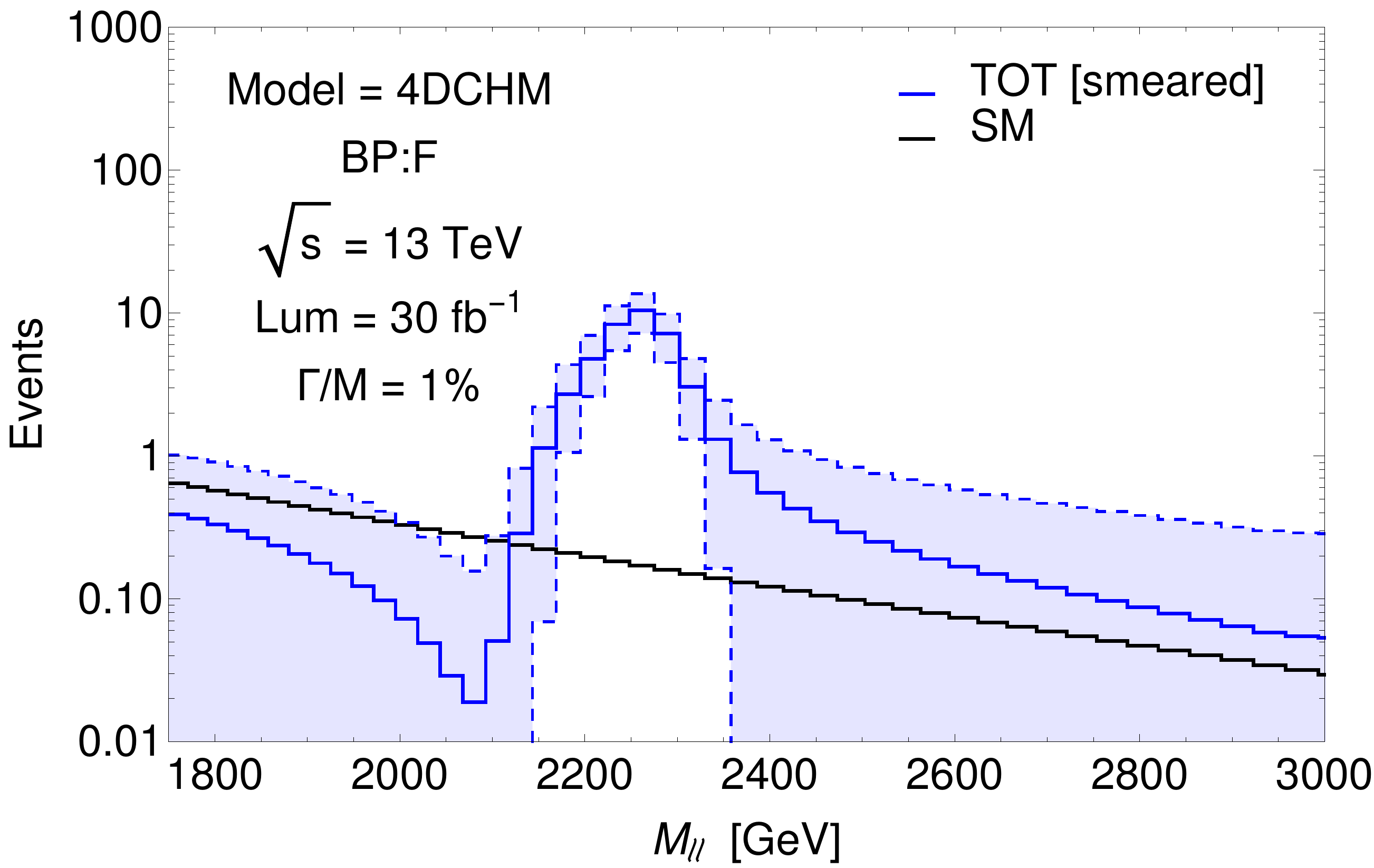}
\includegraphics{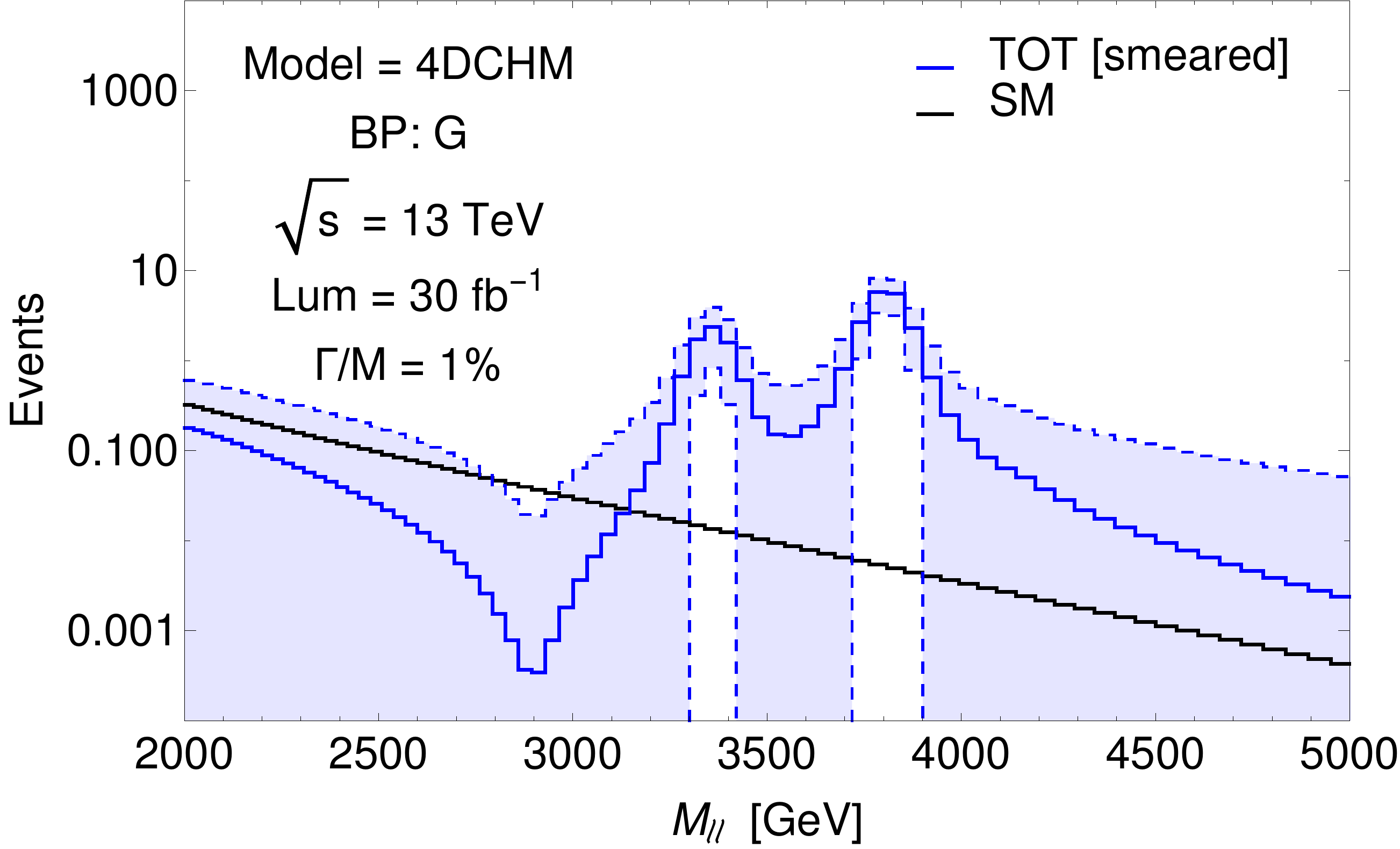} }
 \caption{
 Left(Right): Differential cross section in the dilepton invariant mass for the BP F(G) after the smearing due to finite detector resolution. The width of the gaussian is assumed 26 GeV (38 GeV). The statistical error, visualized as the blue bands, is evaluated for 30 fb$^{-1}$. From \cite{Accomando:2016mvz}.} 
 \label{fig:6}       
 \end{figure}

The two resonances $Z_2$ and $Z_3$ are, in some regions of the parameter space, almost degenerate. Therefore they could appear as a single bump in the dilepton invariant mass after convoluting the signal with a gaussian  distribution to simulate detector resolution. This is shown in Fig.~\ref{fig:6} (Left) for the benchmark F of Table \ref{tab:benchmarks}. 
In Fig.~\ref{fig:6} (Right) the two resonances corresponding to he benchmark G of Table \ref{tab:benchmarks} are separated  enough so that the two peak structure is not washed out after the inclusion of the detector smearing.

\begin{table}
\centering
  \begin{tabular}{|c||c|c|c|c|c|}
    \hline
      Benchmark & $f$ [GeV] & $g_\rho$ & $M_{Z_2}$ [GeV] & $M_{Z_3}$ [GeV] & $M_{Z_5}$ [GeV]\\
    \hline
      F &  1200 & 1.75 & 2192 & 2258 & 2972\\
    \hline
      G & 2900 & 1.00 & 3356 & 3806 & 4107\\
    \hline
  \end{tabular}
  \caption{4DCHM parameter space points associated to the benchmarks F, G  mentioned in the text. From \cite{Accomando:2016mvz}.}
  \label{tab:benchmarks}
\end{table}

 \begin{figure}[h]
 \vspace{-0.5cm}
\centering
\resizebox{0.5\columnwidth}{!}{%
\includegraphics{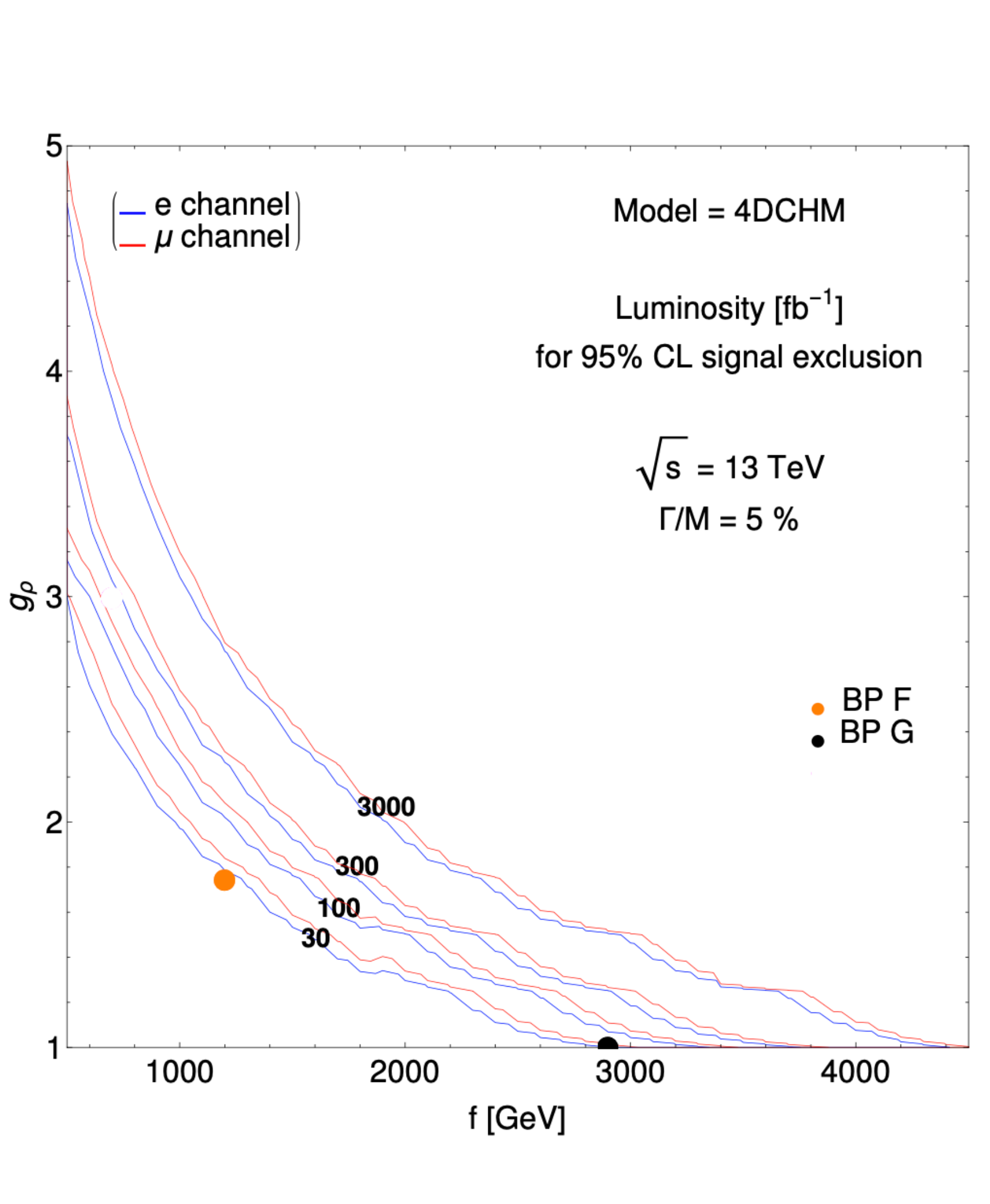} }
 \vspace{-0.3cm}
 \caption{95\%CL exclusion limits from D-Y production in dilepton channel  at the LHC RUNII with 13 TeV for different values of integrated luminosity. The blue (red) contour refers to the electron (muon) channel. From \cite{Accomando:2016mvz}.}
 \label{fig:7}       
 \end{figure}
In Fig.~\ref{fig:7} we show the exclusion limits at LHC RUN II for different values of integrated luminosity (30-3000 fb$^{-1}$) in the parameter space $(f,g_\rho)$. 

The 4DCHM can be considered as a concrete realization of the Model B, described in  subsection (\ref{sec:3_2})  with the feature that the two almost degenerate triplets of spin-1 resonances corresponding to the $SO(4)$ symmetry (broken by the EW interactions) are both phenomenologically relevant. Their interference cannot be neglected as well as their widths which is naturally non-narrow.  A direct comparison of the results shown in Fig.~\ref{fig:7}  with the results shown in Fig.~\ref{fig:1} is not possible, because the analysis of CMS refers to a single heavy vector triplet and takes into account, in addition to the D-Y processes,  diboson  and H-boson productions.

\section{Conclusions}
The presence of new spin-1 particles, heavier than  $W$ and $Z$, is a common feature of several proposals of BSM physics, like for example extensions of the SM in flat or warped extra dimensions, composite Higgs or technicolor models.
In this note we have first reviewed  a general model independent approach based on a simplified effective Lagrangian describing the interactions of a new heavy triplet with the SM particles. This parameterization has been used, using CMS and ATLAS results, to get bounds on the parameters of the effective Lagrangian. We have then reviewed a general approach to built effective Lagrangians describing the interactions in presence of two new triplets of heavy resonances  and considered two particular models and their present limits. Finally we have discussed, in the framework of a composite Higgs model the relevance of the interference and finite width effects, in the Drell-Yan differential cross section  in the dilepton neutral channel. The interference effect, coming from the contributions of  the neutral spin-1 resonances exchanged in the s-channel, generate a  dip before the resonant peak, that spoils the simple analysis performed within the narrow width approximation   or in the finite width approach.  In these cases a model dependent approach is therefore necessary to get bounds on the parameters of the model.


\end{document}